# Research on the Development of Blockchain-based Distributed Intelligent Healthcare Industry: A Policy Analysis Perspective

1st Yang Yue
Institute of Science Studies and
S&T Management
Dalian University of Technology
Dalian, China
ysynemo@mail.edu.dlut.ed.cn

2nd Joseph Z. Shyu
Institute of Science Studies and
S&T Management
Dalian University of Technology
Dalian, China
joseph@cc.nctu.edu.tw

*Abstract*—As a pivotal innovation in digital infrastructure, blockchain ledger technology catalyzes the development of nascent business paradigms and applications globally. Utilizing Rothwell and Zegveld's taxonomy of twelve innovation policy tools, this study offers a nuanced comparison of domestic blockchain policies, dissecting supply, environment, and demand-driven policy dimensions to distill prevailing strategic orientations towards blockchain healthcare adoption. The findings indicate that blockchain technology has seen rapid growth in the healthcare industry. However, a certain misalignment exists between the corporate and policy layers in terms of supply and demand. While companies focus more on technological applications, existing policies are geared towards regulations and governance. Government emphasis lies on legal supervision through environmental policies, aiming to guide the standardization and regulation of blockchain technology. This maintains a balance between encouraging innovation and market and legal regulatory order, thereby providing a reference for the development of the distributed intelligent healthcare industry in our country.

*Keywords—Blockchain, Blockchain Healthcare, Distributed Intelligent, Healthcare Industry, Policy Analysis, Innovation Networks*

## I. Introduction

Influenced by the COVID-19 pandemic, future industrial development strategies and planning have once again become the subject of widespread discussion[1]. The global industrial landscape is shifting from hyper-globalism to hyper-nationalism, with particular emphasis on the critical role of healthcare industries in pandemic management and control[2]. The question of how nations, industries, and enterprises can leverage this global industrial inflection point to create competitive advantage and sustain high-value, high-tech economic growth deserves in-depth consideration.

Owing to advances in precision medicine, the healthcare sector is experiencing rapid growth. In public health management, user-centric healthcare, and the fight against counterfeit drugs, blockchain technology is revolutionizing the industry. The integration blockchain with advances in healthcare and information technology has become a significant trend, especially in healthcare, medical research, and insurance sectors. The healthcare industry faces numerous challenges, notably its complex management system encompassing multiple stakeholders: doctors, researchers, practitioners, support staff, administrators, and patients. Patient data categorization and management pose substantial challenges in data storage and exchange, further complicated by varying data structures and workflows across different healthcare domains. This lack of effective exchange of healthcare-related information between these domains constitutes a significant barrier.

To address these trust-related issues, current researchers and policymakers are turning to blockchain technology. In domestic policy, blockchain technology is included in the Thirteenth Five-Year Plan. China's healthcare system is well prepared for the development of healthcare blockchain; it has a relatively high level of electronic medical data, with 79% in drug management and 13% in patient visit information. With an average of 42 mobile network users per 100 people, the foundational conditions are strong. Blockchain technology in the healthcare industry is being recognized as the next significant opportunity. It is noteworthy that the core competitive advantage of industrial development stems from the support of national policies and regulations, as well as comprehensive planning of industrial structures and enterprise core resources. During the "Fourteenth Five-Year" period, the basic strategy will focus on the "dual circulation" model that leverages China's domestic economic cycle while enhancing it through the Belt and Road Initiative. This aims to counter the impacts of a comprehensive decoupling from the United States and various uncertainties in the post-pandemic era. The further advancement of new infrastructure technologies, including AI, 5G, blockchain, big data, will promote balanced development in China's "Five-in-One" system (economy, society, politics, culture, and ecology)[3].

President Xi Jinping explicitly pointed out in October 2019 that blockchain should be viewed as a crucial technological breakthrough for autonomous innovation, calling for targeted efforts and increased investment to master key technologies and accelerate both blockchain technology and industrial innovation. His remarks provided a clear direction for the future development of China's intelligent industries, as well as a significant strategic blueprint for the future layout of decentralized smart healthcare[4]. Blockchain technology was mentioned for the first time in the State Council's 2016 National Informatization Plan, being included in the conceptual framework for restructuring shared healthcare models.

## II. BLOCKCHAIN APPLICATIONS IN HEALTHCARE REVIEW

The COVID-19 pandemic has exposed the vulnerabilities in existing healthcare systems but also presents significant opportunities for the growth of the healthcare industry [5]. Current healthcare systems, including Clinical Information Systems (CIS), telemedicine, home care services, Integrated Health Information Networks (IHIN), and secondary non-clinical systems, are adopting EHRs for faster access to records. With the continuous increase in medical data, distributed technologies are required for effective management. However, using cloud services necessitates complete trust in third-party providers, both legally and technically. In distributed cloud storage, unencrypted data poses risks to sensitive medical information, especially with the rising threat of cyber-attacks. Hence, blockchain is considered a viable option for secure medical data storage and will revolutionize information-sharing mechanisms [6].

The Potential of Blockchain in Addressing Quality and Safety Issues in Pharmaceutical Supply Chains [7]: This theme encompasses all other codes and elucidates how blockchain resolves challenges posed by cross-border environments and multiple stakeholders, thereby enhancing the authenticity and traceability of pharmaceuticals and reducing legal risks. (1) Quality and Safety: The study links counterfeit drugs, ingredient and dosage issues, and legal disputes together [8]. (2) Supply Chain Complexity: The article connects cross-border settings, handover issues, and multiple stakeholders (producers, carriers, distributors, retailers, and pharmacies together. (3) Blockchain Solutions: The text correlates blockchain characteristics, authenticity and traceability, auditable security records, and role allocation together. In the healthcare sector, blockchain technology is positively impacting the healthcare outcomes for companies and stakeholders, optimizing business processes, improving patient outcomes, enhancing data management, strengthening compliance, reducing costs, and better utilizing healthcare-related data.

Blockchain Applications in Clinical Trials: A Research Study on How Blockchain Technology Addresses Various Issues in Clinical Trials: Specific applications of blockchain in clinical trials include: (1) Complexity in Data Management: This code addresses multiple aspects of data management in clinical trials, such as data storage, transmission, and sharing, as well as the limitations of existing manual systems. (2) Data Integrity and Security: This code focuses on the risks of data being tampered with or erroneously altered, and how blockchain provides solutions. (3) Data Credibility and Verification: This code primarily centers on how, through blockchain technology, the credibility of data can be enhanced, and data verification becomes more secure and effective.

Improvements in Healthcare Information Exchange through Blockchain Applications: The research explores how blockchain technology can bring advancements in the field of healthcare information exchange [9], focusing on the following aspects: (1) Data Interoperability: The emphasis is on how data interoperability is a key factor for the success of healthcare information exchange. (2) Innovation and Decentralization: Discusses how blockchain provides innovative and decentralized solutions. (3) Cost and Efficiency: Highlights how blockchain improves the efficiency of healthcare information exchange by reducing costs and friction. (4) Data Integrity: Focuses on the importance of data integrity in healthcare information exchange. (5) Technological Limitations and Scalability: Analyzes current technological limitations, such as block size, and how these impact the scalability of healthcare information exchange.

Digital Identity Management in Healthcare: Scholars primarily focus on how digital identity management can improve various aspects of healthcare. The paper includes: (1) Security and Privacy: Highlights the shortcomings of traditional centralized systems in terms of security and privacy. (2) Data Integrity: Describes how a distributed model ensures data integrity. (3) User Control and Autonomy: Emphasizes the freedom and control that distributed identity management systems offer to users. (4) Supply Chain Management: Discusses how blockchain can be employed to enhance healthcare supply chain activities [10]. (5) Sensitivity and Confidentiality: Focuses on how blockchain can better protect sensitive and confidential data.

Healthcare Supply Chain Management and Blockchain: Researchers focus on how blockchain profoundly impacts various aspects of the healthcare supply chain. The study includes: (1) Data Integration and Storage: Concentrates on how blockchain can more efficiently manage and store healthcare data. (2) Transparency and Traceability: Describes how blockchain enhances the transparency and traceability of supply chain activities. (3) Drug Flow Monitoring: Explicates how blockchain can be used to curb the circulation of counterfeit drugs. (4) Intermediary Elimination and Streamlining: Points out how blockchain can streamline Health Information Exchange (HIE) systems. (5) Device and IoT Integration: Explores how blockchain and healthcare IoT can be integrated for added benefits. Equally important is blockchain technology's ability to impact the flow of drugs and medical devices in the long and complex healthcare supply chain. The application of blockchain in healthcare supply chains is expected to eliminate the risks of counterfeit drugs that endanger patients globally.

Integrative Healthcare Data Management Using Blockchain Technology [11]: Implementing a secure, efficient, and integrative healthcare data management system. Information Sharing and Data Security [12]: Connecting health information sharing with data sensitivity, as both are core issues in a blockchain-based healthcare system. (1) Data Sensitivity: Describes the characteristics of healthcare information as sensitive data. (2) System Architecture: Introduces multiple system architectures used for implementing data sharing [13]. (3) Technical Details: Elaborates on the technologies and tools used to create these architectures. (4) Data Layers: Mentions the different data layers within the system and their functions. (5) Security and Auditing: Emphasizes how blockchain enhances security and auditing capabilities.

The study explores various approaches, technologies, and regulations involved in patient monitoring using blockchain. First, the primary methods of monitoring involve the use of sensors and blockchain technology. Sensors are employed to monitor the patient's condition, while blockchain is used to secure the generated sensitive personal information. These technologies are not only applied in hospitals and clinics but also encompass the Internet of Things (IoT) and Wireless Body Area Networks (WBAN). Second, various rules and regulations for data protection are an essential component in patient monitoring [14]. Different countries and regions have their specific data protection laws, such as Brazil's LGPD, Europe's GDPR, and the U.S.'s HIPAA, aimed at safeguarding the confidentiality and privacy of patient information. Lastly, emerging technologies and methods, including IoT and WBAN, are expected to further enhance the efficiency and accuracy of patient monitoring. IoT, in particular, offers new possibilities for patient care through networks of sensors and wearable devices [15]. In summary, patient monitoring is a multi-faceted, multi-technological, and multi-regulatory complex system that requires an integration of various factors to achieve effective and secure patient care [16].

Scholars have discussed how blockchain promotes the devolution of authority in the healthcare ecosystem, particularly through the promotion of decentralization for fairness and efficiency. (1) Firstly, decentralization is considered a key factor for achieving fairness and efficiency, a view supported by multiple studies. Beyond structural changes, the economic aspect is also considered, especially concerning maximizing revenue. In the context of increased anti-corruption measures in Chinese healthcare, how to balance the decentralization-induced medical corruption has become a concern for policymakers. (2) Secondly, transparency in data transactions is deemed an important issue. Transparency not only increases the system's credibility, but also ensures, to an extent, customer rights and data security [17]. Customer engagement, particularly through the establishment of a fair customer role, further amplifies the system's fairness and transparency. Through multi-dimensional analyses, the multiple benefits of blockchain in healthcare are demonstrated, especially in promoting the devolution of authority. These advantages extend beyond structural and economic dimensions to include transparency and customer engagement, offering rich insights for further research and practice [18]. The cumulative effects indicate that blockchain technology has extensive and far-reaching impacts, warranting further exploration and application.

Through the integration of platforms, optimization of electronic health systems, and expansion of service scope, blockchain technology can not only enhance the quality and efficiency of medical services but also offer more possibilities for future healthcare. Scholars have explored how to construct a smart healthcare ecosystem through blockchain technology. This is primarily achieved through three dimensions: the application of blockchain, optimization of electronic health systems, and the expansion of service scope [19]. (1) First, integrating blockchain platforms into the healthcare ecosystem can create intelligent healthcare systems. Such integration not only enhances system interoperability but also strengthens collaboration among stakeholders. (2) Second, the potential of blockchain in optimizing electronic health systems. By using blockchain, a more efficient and secure electronic health record system can be built. (3) The development of blockchain-based electronic health and telemedicine information systems will potentially expand the service scope of healthcare providers. Such expansion not only improves service coverage but also accommodates the diverse needs of different patients.

## III. RESEARCH METHODS

Industry-specific trajectories necessitate adaptive strategies, demanding precision in governmental and corporate planning to align with phase-specific and environmental exigencies. Employing Rothwell and Zegveld's framework, we dissect the multifaceted components underpinning industrial innovation (refer to Figure 1). Such elements call for a nuanced analysis and investigation into the innovation infrastructure. Through policy adjustments, nations and enterprises can alter relevant conditions and factors, thereby gaining competitive advantage. Technological advancements and innovation are at the core of emerging theories of competitive advantage. The government primarily plays a role here by promoting innovative technologies. In the absence of governmental intervention, firms often underperform in technological investments. Consequently, government entities bear the onus of furnishing targeted backing and inducements to bolster development initiatives.

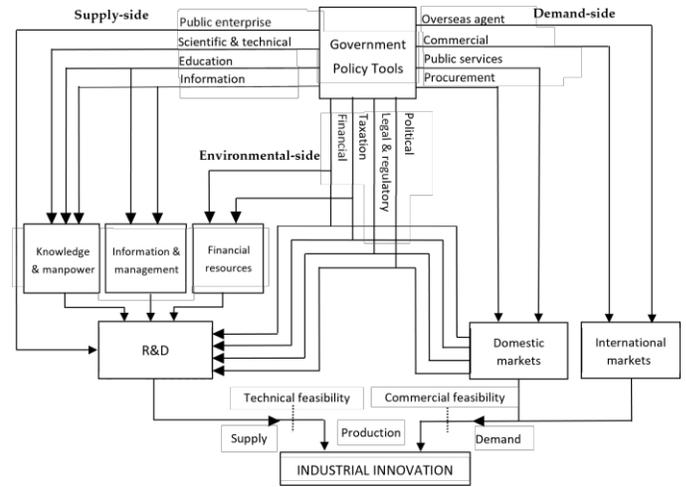

Fig. 1. Inducing innovation – the required factors of the industrial innovation. Source: Rothwell & Zegveld .

For governments, successful innovation strategies necessitate the establishment of apt linkages between technological supply and market demand. Rothwell and Zegveld categorize policy instruments into 12 types: public enterprises, science and technology development, education and training, information services, financial services, tax incentives, legal regulations, strategic policies, government procurement, public services, trade controls, and overseas institutions. From an industrial perspective, a government's innovation policy should encompass both technological and industrial strategies. Government activities are composed of 12 industry innovation policy tools, which are further classified into 'supply-side' policies, 'environmental' policies, and 'demand-side' policies. Successful innovation aligns appropriately with supply-side technology and demand-side market factors. Policy tools of blockchain eHealth industry

TABLE I. ROTHWELL AND ZEGVELD POLICY INSTRUMENTS

| Dimension | Policy tools | Remarks |
| --- | --- | --- |
| **Supply Side Policy** | (1) Public enterprise | Involvement in Public Utilities by Government: Includes Establishment, Operation, and Management |
| | (2) Scientific and technical | Government's Direct or Indirect Encouragement of Science and Technology Development |
| | (3) Education | Policies on Educational Systems and Vocational Qualifications by Government |
| | (4) Information | Government's Direct or Indirect Facilitation of Technology and Market Information Flow |
| **Environmental Side Policy** | (5) Financial | Government's Direct or Indirect Financial Support to Enterprises |
| | (6) Taxation | Government Tax Incentives for Enterprises |
| | (7) Legal and regulatory | Special Measures to Regulate Market Order by Government |
| | (8) Strategic Industries | Strategic Measures for Industrial Development Provided by Government |
| **Demand Side Policy** | (9) Public services | Service Measures for Social Issues |
| | (10) Commercial | Regulations on Import and Export |
| | (11) Overseas agent | Government's Direct or Indirect Support for Overseas Branch Establishment |

Content analysis systematically presents the blockchain policy tools employed. The analysis of the data set provides a superior interpretation of innovative blockchain strategies. Key attributes of blockchain strategies are described through quantitative explanations derived from descriptive statistics. First, we gather relevant policy texts on blockchain from our national government documents, focusing on blockchain policy texts published in China. Second, we employ Rothwell and Zegveld's twelve categories of government policy instruments and transform China's blockchain policy texts into government innovation activities, as shown in Figure 1. Based on the classification results of government activities, we calculate the proportions of each of the twelve policy tools in the related policies of our country. Third, we incorporate the results of our analysis of China's blockchain innovation policy framework into the innovation network model to identify the critical success factors that need development.

## IV. FINDING

### A. Analysis of China's Blockchain Healthcare Policy

The study comprehensively analyzes the adoption of blockchain policies across 22 provinces in China, utilizing data from diverse sources, including government publications and reports. The results were visually represented using a vertical bar chart for ease of comparison. The findings indicate that the average number of blockchain policies per province stands at 4.36, with a standard deviation of 3.91. The provinces with the most active adoption of blockchain policies are Guangdong (16 policies), Guizhou (11 policies), Zhejiang (10 policies), Shanghai (8 policies), and Beijing (7 policies). Conversely, provinces such as Shaanxi, Liaoning, Hubei, Inner Mongolia, and Xinjiang lag behind with only one policy each. This disparity reveals that there is significant room for growth in the adoption of blockchain policies in certain regions.

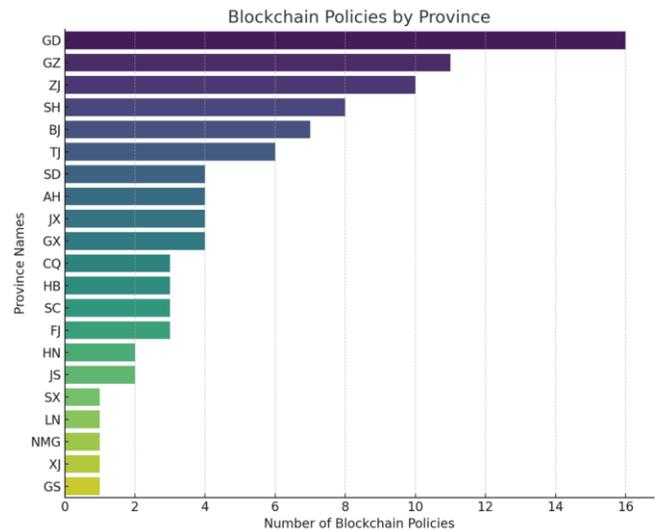

Fig. 2. Blockchain Policy Distribution in China

Utilizing the twelve policy tools, this study analyzes China's national-level blockchain policies. This analysis aims to understand the different emphases in supply, environmental, and demand aspects of China's recent blockchain policies. The analysis results are illustrated in Figure 3. Regulations and governance, along with industrial strategy, have the most significant weight, followed by education and training. This indicates that in the face of the nascent state of blockchain technology in the industry, there is an urgent need for the Chinese government to establish technical standards for blockchain. Building Key Success Factors for the Healthcare Blockchain Innovation Network: An analysis is conducted on the critical elements that contribute to the success of our country's healthcare blockchain industry. The corresponding supply-side, environmental, and demand-side factors within the 12 policy tools are identified. Comparative Analysis: A contrasting study is undertaken to identify differences in emphasis on blockchain implementation from both an industrial and policy perspective.

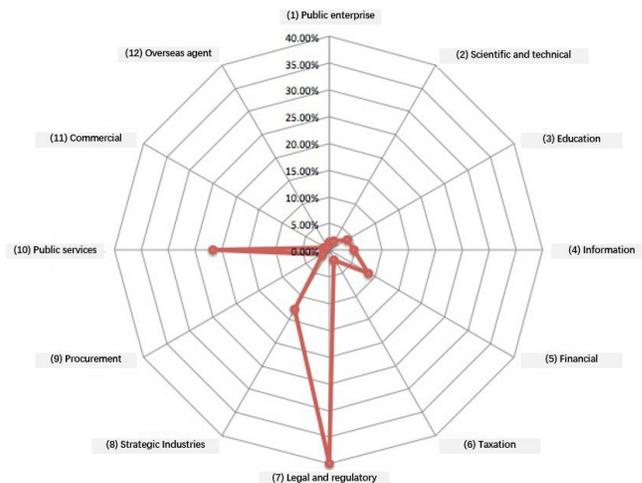

Fig. 3. Radar chart of blockchain policies in China

This paper analyzes the key success factors for enterprises in the blockchain medical field through an innovation network model, and evaluates the effectiveness of corresponding policy tools in promoting blockchain technology development in China. The key takeaways are as follows:

(1) There are significant gaps in knowledge platform economics, technical leadership, state-monopoly management, and large-scale customized business models, indicating that China's blockchain medical industry has a long way to go. The focus currently lies in hardware manufacturing, platform services, security services, talent services, and training.

Trust and security are notably the most compelling attributes of transitioning from smart healthcare to distributed smart healthcare based on blockchain. Blockchain in healthcare plays a significant role in addressing various challenges in the medical field. It offers secure data storage and sharing services to medical institutions, promotes pharmaceutical R&D capabilities, and fosters a tiered diagnosis and treatment system.

Efforts should be made to integrate artificial intelligence and blockchain in smart healthcare, starting with interoperability without reconciliation and cybersecurity. Blockchain should be considered a critical breakthrough in technological innovation, and increased investments should be made to conquer a series of key core technologies to accelerate the innovative development of blockchain technology and the healthcare industry. Strengthen basic research and improve original innovation capabilities, aiming for China to lead in theory and innovation and achieve new industrial advantages in this emerging field. Promote coordinated R&D efforts to speed up breakthroughs in core technology and provide secure, controllable technological support for blockchain application development.

Seize opportunities in technology integration, functional expansion, and industrial segmentation to leverage blockchain's role in promoting data sharing, optimizing business processes, reducing operational costs, improving collaboration efficiency, and building trustworthy systems. Strengthen guidelines and standards for blockchain technology, and research and analyze the security risks associated with blockchain to explore the establishment of a security system adapted to blockchain technology mechanisms.

(2) Textual Analysis of 12 Policy Tools on China's Blockchain Policy, our analysis of China's 12 policy tools focused on blockchain reveals that the emphasis lies on the supply and environmental aspects, with the highest policy weightage in education and training, legal regulations, and strategic planning. This reflects the government's desire to actively guide the blockchain medical industry, encouraging its development from a top-down approach and promoting autonomous growth within a market framework. Forward deployment from a top-down design is critical for the smart healthcare industry, aimed at solving the sector's challenges and establishing a distributed economic governance model for the blockchain in smart healthcare. Legal regulations, strategic policy-making, and the integration of public services with the healthcare industry will play mutually beneficial roles.

In the absence of government intervention, businesses typically underinvest in technology. Moreover, industries have varying needs during different developmental phases and in different environments. Hence, the government and industry could make appropriate plans based on pinpointing resource needs throughout the development process. Both state and enterprises can alter related factors and conditions via policy to gain competitive advantages. Requirements for resources will differ based on time and environmental factors. Further comparative analysis by Porter (1990) suggests that, in addition to elevating competition to the national level, technological advancement and innovation must be focal points for consideration.

*B. Analysis of Healthcare Blockchain Innovation Network*

An Innovative network analysis as follows: In the Innovation Space, "Defined" refers to issues that are well-defined or structured, focusing on development, utilization, and efficiency. It emphasizes the application of existing knowledge or assets to establish links based on existing knowledge foundations. "Arbitrary" refers to issues that are hard to define or structure, necessitating exploration, creativity, and innovation. It creates new knowledge or assets and reflects unknown connections in the knowledge system. In Network Leadership, "Centralized" leadership is led by participants within the domain, featuring a formal organizational structure that embodies a hierarchical architecture and a closed decision-making process, with a clear core-peripheral community. "Decentralized" is community-led, presenting a distributed organizational structure with an open decision-making mode and an unclear community core. Textual analysis of the collected materials and questionnaire design contribute to constructing a healthcare blockchain innovation network model for our country, as shown in Figure 4. This model is divided into Knowledge Platform Economic Models, Professional Technical Leadership Models, Oligopoly State Operating Models, and Large-Scale Customized Operating Models. Each includes specific success factors for the healthcare blockchain industry.

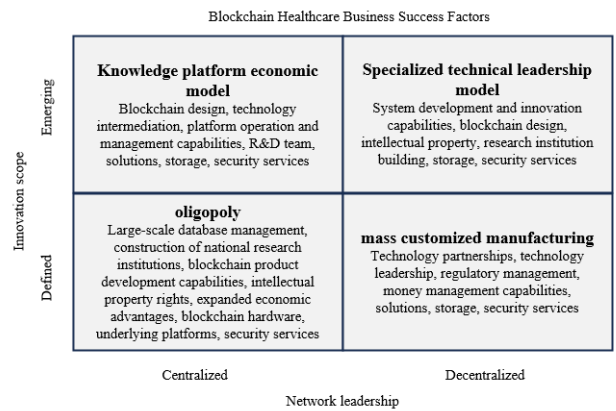

Fig. 4. Blockchain Healthcare Business Success Factors

This paper primarily employs the Network Innovation analysis method to delve into the healthcare blockchain sector in our country and to evaluate the policy support it requires. Nambisan and Sawhney define Network-Centered Innovation (NCI) as "an externally-focused innovation approach that relies on leveraging external networks and communities to expand or enhance the scope, speed, and quality of innovation," as depicted in Figure 3.Blockchain. The innovation network model identifies key success factors that need development. In the construction aspect, the healthcare blockchain industry requires a high degree of emphasis on elements such as hardware manufacturing, platform services, security services, talent services, and training.

## V. CONCLUSIONS

China's blockchain industry is in its nascent stage, with an emerging industrial ecosystem and a gradually constructed policy framework. The environment for industrial development is continuously improving, and the speed of enterprise development and their numbers are accelerating, with a relatively concentrated regional distribution. There is a clear agglomeration effect in the industry, and the structure of the blockchain industrial chain is becoming increasingly defined, continually supported by funding. This paper analyzes key documents related to blockchain policy issued by the government. Our findings from policy analysis reveal that our country is actively establishing scientific and technological infrastructure for blockchain in healthcare, accompanied by various legislative and incentive measures to encourage academia and industry in their efforts towards R&D technology adoption and dissemination. This provides valuable insights into the current positioning and future trajectory of this emerging technology sector.

The study synthesizes the diverse foci of our country's blockchain sector at the industrial and policy levels. Policy formation primarily hinges on the combination of these policy instruments, which functionally fall into fiscal, human, and technological support categories. These tools play a role in providing resources for technological innovation and the production process. Moreover, government actions like technological contract research and public procurement impact innovation and marketing processes, serving as tools for creating market demand. Additionally, establishing infrastructure for technological advancement, along with various incentives and regulatory measures, encourage efforts in R&D, technology adoption, and dissemination from academia and industry.

These documents encompass the government's regulatory framework and strategic roadmap for blockchain in healthcare, representative of future trends in healthcare like smart medical devices, customized healthcare, and preventive medicine. The government must formulate a series of innovative policies to guide, safeguard, and promote the development of the domestic smart healthcare industry, providing policy support for the rapid growth of the blockchain. In the context of our country's burgeoning smart healthcare industry, current focus mainly lies in hardware manufacturing, platform services, security services, talent services, and development—all of which are critical to the growth of the healthcare blockchain sector. Strengthening regulatory measures is crucial for enhancing the overall operational efficiency of the healthcare industry, fostering innovative medical services, and establishing mechanisms. These mechanisms should encourage hospitals to transition from being mere managers of medical data to active providers of blockchain services, thereby actively participating in the building of public blockchain infrastructure. This will facilitate a shift from isolated, profit-driven operations to a more collaborative and stratified healthcare system. Given the continuous development of blockchain technology, enterprises must also consider corresponding technical standards, legal regulations, and talent cultivation, while simultaneously supporting the development of specialized talent in blockchain."

REFERENCES

[1] Huang R, Chen Wei z, et al. Optimization Research of the National Infectious Disease Monitoring and Early Warning Network Based on Blockchain Technology[J]. Chinese Journal of Management, 2020,17(12):1848-1856.

[2] Fusco A, Dicuonzo G, Dell'Atti V, et al. Blockchain in healthcare: Insights on COVID-19[J]. International Journal of Environmental Research and Public Health, 2020, 17(19): 7167.

[3] Yue, Yang, and Joseph Z. Shyu. 2023. "An overview of research on human-centered design in the development of artificial general intelligence." arXiv preprint arXiv:2309.12352. https://doi.org/10.48550/arXiv.2309.12352.

[4] 杨岳 王贤文.(2021).Altmetrics 十年发展综述.情报杂志(11),

[5] Arumugam S K, Sharma A M. Role of Blockchain in the Healthcare Sector: Challenges, Opportunities and Its Uses in Covid-19 Pandemic[C]//International Conference on Hybrid Intelligent Systems. Springer, Cham, 2021: 657-666.

[6] Avdoshin S, Pesotskaya E. Blockchain revolution in the healthcare industry[C]//Proceedings of the Future Technologies Conference. Springer, Cham, 2018: 626-639.

[7] Patel V. A framework for secure and decentralized sharing of medical imaging data via blockchain consensus[J]. Health informatics journal, 2019, 25(4): 1398-1411.

[8] Sanmarchi F, Toscano F, Fattorini M, et al. Distributed solutions for a reliable data-driven transformation of healthcare management and research[J]. Frontiers in Public Health, 2021, 9: 944.

[9] Abunadi I, Kumar R L. BSF-EHR: blockchain security framework for electronic health records of patients[J]. Sensors, 2021, 21(8): 2865.

[10] Omar I A, Jayaraman R, Debe M S, et al. Automating procurement contracts in the healthcare supply chain using blockchain smart contracts[J]. IEEE Access, 2021, 9: 37397-37409.

[11] Khatoon A. A blockchain-based smart contract system for healthcare management[J]. Electronics, 2020, 9(1): 94.

[12] Y. Yue and J. Z. Shyu, Reinventing Library Knowledge Services through Librchain: An Open Innovation approach[J/OL], https://arxiv.org/submit/5087123/view

[13] Al Omar A, Bhuiyan M Z A, Basu A, et al. Privacy-friendly platform for healthcare data in cloud based on blockchain environment[J]. Future generation computer systems, 2019, 95: 511-521.

[14] Ismail L, Zeadally S. Healthcare insurance frauds: Taxonomy and blockchain-based detection framework (Block-HI)[J]. IT professional, 2021, 23(4): 36-43.

[15] Yue, Yang, and Joseph Z. Shyu. "Blockchain-Based Open Network in Technology Intermediation." In 2019 IEEE International Symposium on Innovation and Entrepreneurship (TEMS-ISIE), pp. 1-7. IEEE, 2019.

[16] Subramanian G, Thampy A S. Implementation of blockchain consortium to prioritize diabetes patients' healthcare in pandemic situations[J]. Ieee Access, 2021, 9: 162459-162475.

[17] Zerga H, Amraoui A, Benmammar B. Distributed, dynamic and trustworthy access control for telehealth systems[J]. Concurrency and Computation: Practice and Experience, e7352.

[18] Fu J, Wang N, Cai Y. Privacy-preserving in healthcare blockchain systems based on lightweight message sharing[J]. Sensors, 2020, 20(7): 1898.

[19] Mendes D, Rodrigues I, Fonseca C, et al. Anonymized distributed PHR using blockchain for openness and non-repudiation guarantee[C]//International Conference on Theory and Practice of Digital Lib